\documentclass{article}

\usepackage{arxiv}

\usepackage{amssymb}
\usepackage{latexsym}

\usepackage{url}
\usepackage{xcolor}

\usepackage{hyperref}

\usepackage{booktabs, caption, makecell}
\usepackage{adjustbox}
\usepackage{threeparttable}
\usepackage{amsmath}
\usepackage{soul}
\usepackage{calrsfs}

\definecolor{newcolor}{rgb}{.8,.349,.1}

\usepackage[utf8]{inputenc} % allow utf-8 input
\usepackage[T1]{fontenc}    % use 8-bit T1 fonts
\usepackage{hyperref}       % hyperlinks
\usepackage{url}            % simple URL typesetting
\usepackage{booktabs}       % professional-quality tables
\usepackage{amsfonts}       % blackboard math symbols
\usepackage{nicefrac}       % compact symbols for 1/2, etc.
\usepackage{microtype}      % microtypography
\usepackage{lipsum}
\usepackage{graphicx}
\title{Upper-body free-breathing Magnetic Resonance Fingerprinting applied to the quantification of water T1 and fat fraction}%

\author{
 Constantin Slioussarenko \\
  Institute of Myology, Neuromuscular Investigation Center, NMR Laboratory\\
  G.H. Pitié-Salpêtrière\\
  75651 Paris Cedex 13, France \\
  \texttt{c.slioussarenko@institut-myologie.org} \\
   \And
 Pierre-Yves Baudin \\
  Institute of Myology, Neuromuscular Investigation Center, NMR Laboratory\\
  G.H. Pitié-Salpêtrière\\
  75651 Paris Cedex 13, France \\
  \And
 Marc Lapert \\
 Siemens Healthcare SAS\\
 Courbevoie, France\\
  \And
 Benjamin Marty \\
  Institute of Myology, Neuromuscular Investigation Center, NMR Laboratory\\
  G.H. Pitié-Salpêtrière\\
  75651 Paris Cedex 13, France \\
}

\begin{document}
\maketitle
\begin{abstract}
Over the past decade, Magnetic Resonance Fingerprinting (MRF) has emerged as an efficient paradigm for the rapid and simultaneous quantification of multiple MRI parameters, including fat fraction (FF), water T1 ($T1_{H2O}$), water T2 ($T2_{H2O}$), and fat T1 ($T1_{fat}$). These parameters serve as promising imaging biomarkers in various anatomical targets such as the heart, liver, and skeletal muscles. However, measuring these parameters in the upper body poses challenges due to physiological motion, particularly respiratory motion. In this work, we propose a novel approach, motion-corrected (MoCo) MRF T1-FF, which estimates the motion field using an optimized preliminary motion scan and uses it to correct the MRF acquisition data before dictionary search for reconstructing motion-corrected FF and $T1_{H2O}$ parametric maps of the upper-body region. We validated this framework using an \textit{in vivo} dataset comprising ten healthy volunteers (6 men, 4 women, mean age = 39 $\pm$ 12 years old) and a 10-year-old boy with Duchenne muscular dystrophy. At the ROI level, in regions minimally affected by motion, no significant bias was observed between the uncorrected and MoCo reconstructions for FF (mean difference of -0.7\%) and $T1_{H2O}$ (-4.9 ms) values. Moreover, MoCo MRF T1-FF significantly reduced the standard deviations of distributions assessed in these regions, indicating improved precision. Notably, in regions heavily affected by motion, such as respiratory muscles, liver, and kidneys, the MRF parametric maps exhibited a marked reduction in motion blurring and streaking artifacts after motion correction. Furthermore, the diaphragm was consistently discernible on parametric maps after motion correction. This approach lays the groundwork for the joint 3D quantification of FF and $T1_{H2O}$ in regions that are rarely studied, such as the respiratory muscles, particularly the intercostal muscles and diaphragm.
\end{abstract}

%% main text
\section{Introduction}
\label{Intro}

MR fingerprinting (MRF) represents an efficient paradigm introduced a decade ago for rapid and simultaneous quantification of multiple MRI parameters (\cite{maMagneticResonanceFingerprinting2013}). The fundamental concept of MRF involves acquiring a series of highly undersampled images, followed by pixel-wise fitting to a pre-computed dictionary of elements (fingerprints) to extract the underlying tissue parameters. Among the numerous variables measurable by MRF, fat fraction (FF), water T1 ($T1_{H2O}$), water T2 ($T2_{H2O}$) and/or fat T1 ($T1_{fat}$) have demonstrated promise as imaging biomarkers in various anatomical targets, including heart (\cite{jaubertWaterFatDixonCardiac2020,jaubertT1T2Fat2021}), liver (\cite{jaubertMultiparametricLiverTissue2020,velascoSimultaneousComprehensiveLiver2022}) and skeletal muscles (\cite{cenciniMagneticResonanceFingerprinting2019,martyMRFingerprintingWater2020,ostensonMRFingerprintingSimultaneous2019}). For example, the MRF T1-FF sequence introduced by \cite{martyMRFingerprintingWater2020} has been proposed for quantitative monitoring of $T1_{H2O}$ and FF of the skeletal muscle tissues. This approach has been applied at 3T in the lower limb muscles of subjects with different neuromuscular diseases, and allowed the joint reconstruction of $T1_{H2O}$, $T1_{fat}$, FF, B0 and B1 maps (\cite{martyWaterfatSeparationMR2021,gerhalterQuantitative1H23Na2021,fromesGrowingHeartCongenital2023}. However measuring those parameters in the upper body presents challenges due to physiological motion, and more specifically respiratory motion. \\
\indent In particular, respiratory muscles, like the intercostal muscles and diaphragm, are thin structures commonly affected in neuromuscular disorders. Patients often suffer from significant discomfort and sometimes require ventilation support. While qualitative imaging techniques such as echography or fat-saturated T2-weighted MRI (\cite{vieirasantanaDiaphragmaticUltrasoundReview2020,ciceroMagneticResonanceImaging2020}), as well as functional imaging like MR spirometry (\cite{harlaarDiaphragmaticDysfunctionNeuromuscular2022}) have been employed to study these key targets, quantitative MRI, which could potentially identify structural abnormalities at an early stage before functional impairments manifest, has not been conducted due to the challenges posed by breathing motion.\\
\indent In qualitative Cartesian MRI, breathing motion often results in blurring and ghosting artifacts along the phase encoding direction (\cite{woodMRImageArtifacts1985}). In non-Cartesian MRI, while ghosting artifacts may decrease as the center of k-space is sampled multiple times (\cite{gloverProjectionReconstructionTechniques1992}), blurring persists. Early observations indicated a significant impact of motion-induced artefacts on quantitative MRI, resulting in biased parameter estimation (\cite{giriMyocardialMappingRespiratory2012}). \\
\indent Consequently, various methods have been proposed to mitigate breathing motion effects in MRI. Generally, an initial step involves detecting motion during acquisition to split the data into several respiratory phases. This can be achieved through various means. For instance, respiratory bellows, which measure the abdominal cavity expansion and contraction, can be positioned around the patient's abdomen (\cite{ehmanMagneticResonanceImaging1984}). However, this technique may inaccurately detect motion depending on the bellows placement and the patient's breathing strength (\cite{santelliRespiratoryBellowsRevisited2011}). A recently developed device, the Pilot Tone, addresses this issue by emitting a radiofrequency wave at a fixed frequency beyond the MRI acquisition bandwidth (\cite{solomonFreeBreathingRadial2021}). This wave is modulated by various motions, and primarily breathing motion. The advantage of Pilot Tone lies in its flexible placement within the magnet bore, with reduced reliance on patient. Another approach involves using gating spokes (or navigators) integrated within the MRI sequence, acquiring a low-resolution image of the body along a specific direction at precise time intervals (\cite{ehmanAdaptiveTechniqueHighdefinition1989}). The motion of anatomical features such as the lung-liver interface can then be tracked over time to sort the data into the different respiratory phases. This latter method is inherently synchronized with the MRI sequence and eliminates the need for external devices. \\
\indent Once the acquired data has been binned into the different respiratory phases, several strategies can be employed to correct for motion. A first approach involves retaining data from one single respiratory phase, typically full-expiration, which often lasts longer than other phases (\cite{huangFreeBreathingAbdominal2021}). However, this method discards a significant amount of data, potentially leading to undersampling issues. One prospective solution involves extending the MRI acquisition duration to gather sufficient data for reconstructing the desired respiratory phase (\cite{ehmanMagneticResonanceImaging1984}). Nonetheless, this method inevitably leads to longer scan times. In order to use data from all respiratory phases, retrospective motion correction (MoCo) has been proposed. Some methods address MoCo directly in k-space, such as autofocus, which aims to minimize image gradient entropy by applying a phase correction (\cite{atkinsonAutofocusAlgorithmAutomatic1997}). However, this approach is mainly suitable for correcting rigid deformations or piecewise rigid deformations by individually correcting each coil's k-space  (\cite{chengNonrigidMotionCorrection2012}). For free-form deformation, the preferred method is to do it in image space. One approach consists in reconstructing the images for all motion states simultaneously by including low-rank constraints, such as multitasking (\cite{christodoulouMagneticResonanceMultitasking2018}). These methods offer the advantage of not requiring motion field estimation between the different states. However, the minimization problem dimension may be large and is potentially ill-posed, rendering it hardly suitable for high-resolution 3D imaging. Alternatively, one may first estimate the deformation field on low-resolution images and then apply this field to register all images to the same reference. Methods like MoCo with high-dimensional total variation (MoCo-HDTV) (\cite{rank4DRespiratoryMotion2017}) have been proposed in qualitative MRI to reconstruct motion-corrupted images by iteratively estimating deformation fields and reconstructing images, gradually increasing spatial resolution after each step. For MRF at 1.5T, \cite{cruzGeneralizedLowRank2022} proposed estimating deformation fields using denoised singular volumes from various motion states. This deformation field was then incorporated into the least square reconstruction of singular volumes to yield MRF maps after pattern matching. However, larger B1 variations throughout the field of view at 3T, coupled with increased undersampling artifacts resulting from the radial acquisition scheme of the MRF T1-FF sequence, represent a challenge in obtaining motion-resolved singular volumes of adequate quality for image registration.\\
\indent In this study, we propose a novel approach, MoCo MRF T1-FF, to iteratively estimate the motion field from an optimized preliminary motion scan and apply it to correct the MRF acquisition data before pattern matching for reconstructing motion-corrected parametric maps of the upper-body region. From a methodological standpoint, the proposed method offers two significant contributions: by using a preliminary motion scan with separately optimized acquisition parameters for contrast, the deformation field can be precisely estimated, rendering the approach modular. Consequently, the MRF scan can be optimized solely for quantitative parameter estimation without concern for deformation field estimation. From an application standpoint, this study paves the way for joint quantification of FF and $T1_{H2O}$ in 3D in regions rarely studied, such as the respiration muscles, particularly the intercostal muscles and diaphragm.  

\section{Methods}
\label{Methods}

\subsection{MR acquisition framework}
The MR acquisition framework comprised two separate scans: firstly, a 3D radial stack-of-stars Fast Low Angle Shot (FLASH) sequence, referred to as the "motion scan", was used for estimating the deformation fields between the different respiratory phases. This was followed by the 3D MRF T1-FF sequence, as proposed in \cite{martyQuantitativeSkeletalMuscle2021} an acquisition specifically designed for joint estimation of $T1_{H2O}$, FF, B0, B1 and $T1_{Fat}$, to monitor muscle tissue alterations in subjects with neuromuscular diseases. The MRF T1-FF sequence consists of 1400 golden angle radial spokes, incorporating varying echo times and flip angles allowing the reconstruction of a time-series of 175 undersampled images at different contrasts (\cite{martyMRFingerprintingWater2020}). The 3D MRF T1-FF sequence follows this principle, acquiring partitions sequentially to form a time-series of 175 3D volumes using a golden angle stack-of-stars radial sampling scheme (\cite{fengGoldenAngleRadialMRI2022}). It can also be undersampled in the partition direction to accelerate the acquisition (\cite{martyQuantitativeSkeletalMuscle2021}).\\
\indent Navigators were inserted every 28 spokes within both sequences to track the lung/liver interface during acquisition with a sampling frequency of approximately 10 Hz. These 1D gating spokes were centered at a TE of 2.39 ms, to put water and fat signals in phase, thereby maximizing the liver signal and thus the contrast between the lung and liver. These navigators were acquired in the head-foot direction, with the readout center positioned around the dome of the liver and a field of view of 25 cm. A slice-selective RF pulse was applied with a slice thickness of 3 cm to obtain high signal-to-noise ratio and sufficient contrast by minimizing the amount of subcutaneous fat projected onto the navigator image. This parameter was empirically optimized by visually analyzing the sharpness of the lung/liver interface on the navigator images as a function of slice thickness in healthy volunteers.

\subsection{MR reconstruction framework} \label{subsec:MRFreco}
\begin{figure*}[!htbp]
\caption{Diagram illustrating the different steps of the proposed reconstruction framework for reconstructing the motion-corrected parametric maps: 1) Liver/lung interface displacement extraction and binning on the motion scan (Section \ref{subsec:MoCoDetection}); 2) Iterative free-form deformation estimation on the motion scan volumes for each respiratory phase (Section \ref{subsec:MoCoEstim}); 3) Liver/lung interface displacement extraction and binning on the MRF scan (Section \ref{subsec:MoCoDetection}); 4) Iterative reconstruction of MRF singular volumes using the deformation fields estimated in step 2 (Section \ref{subsec:MoCoCorrection}); 5) MRF pattern matching using bi-component dictionary matching, as proposed in \cite{slioussarenkoBiComponentDictionary2024}, to reconstruct $T1_{H2O}$, FF, B0, B1 and $T1_{Fat}$  parametric maps from the motion-corrected singular volumes.}
\label{fig:MoCoFramework}
\centering
\includegraphics[width=\textwidth]{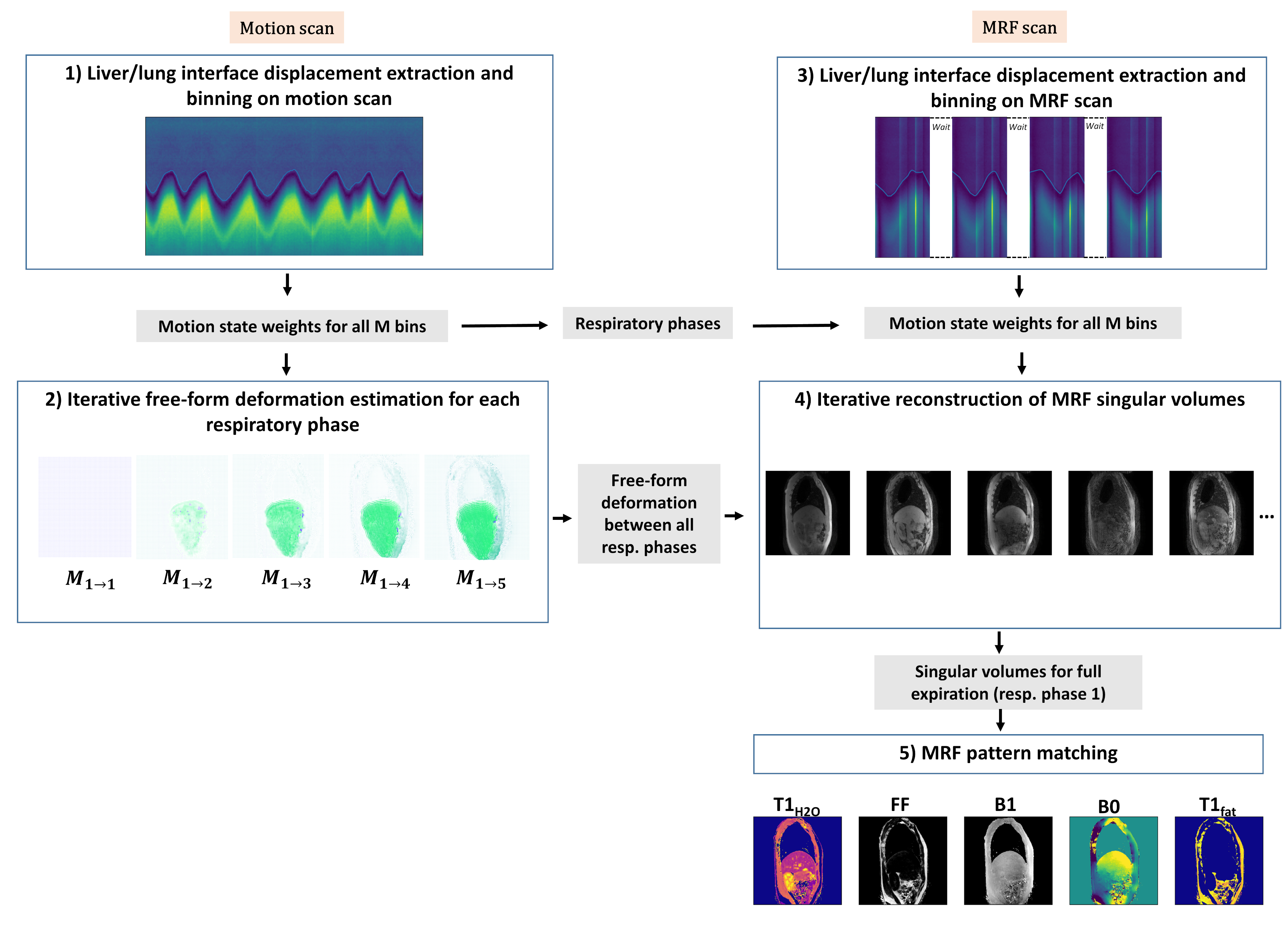}
\end{figure*}

\indent The reconstruction procedure applied both to the motion scan and to the 3D MRF T1-FF sequence to obtain the MoCo parametric maps is summarized in Fig.\ref{fig:MoCoFramework}.

\subsubsection{Displacement extraction and signal binning} \label{subsec:MoCoDetection}
%For displacement calculation, the first step was to select the channel with the most contrast between the lung and liver interface on the gating spokes image of the motion scan. The same channel was used for the displacement calculation on the MRF scan to ensure consistency with the motion scan.
The mean 1D navigator of the motion scan was calculated by averaging the navigator images across all repetitions of the sequence. The displacement $d(n)$ of the liver/lung interface for a specific navigator spoke $n$ was calculated as the spatial shift maximizing the correlation between the shifted navigator spoke and the mean navigator.

\begin{figure*}[!htbp]
\centering
\includegraphics[width=\textwidth]{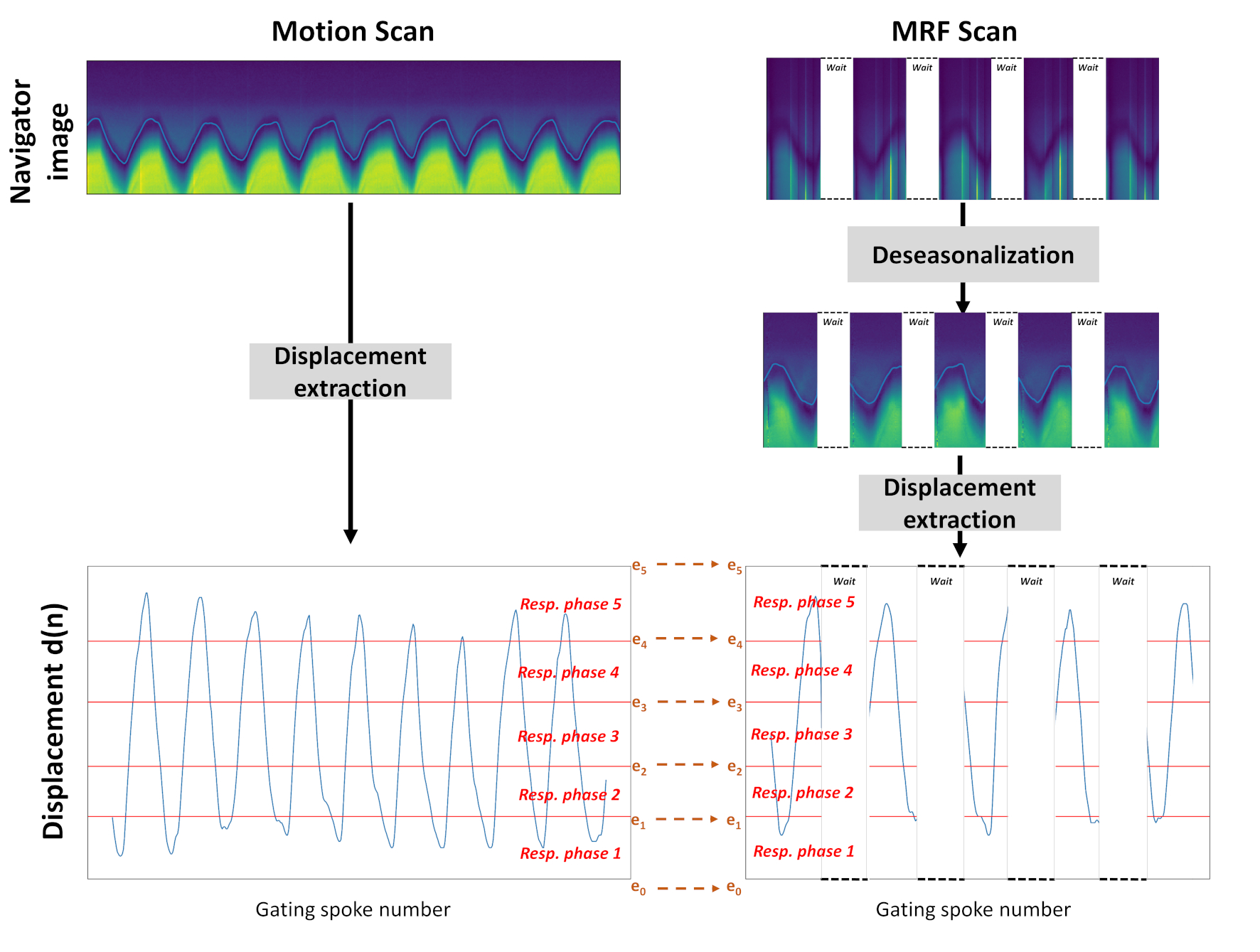}
\caption{Representative examples of the liver/lung interface displacement estimation d(n), along with data binning, derived from both motion and 3D MRF T1-FF scans using navigator images. The navigator images for the 3D MRF-T1-FF scan undergo a deseasonalization procedure initially to eliminate the seasonal MRF component. The z-coordinates of each respiratory phases ($[e_{0},...,e_{M}]$) are determined from the motion scan data to ensure an equal number of spokes per respiratory phase. These coordinates were then applied to bin the MRF scan data. The jumps on the MRF navigator images are due to the fact that there is a recovery time at the end of each repetition to allow for the magnetization to grow back to equilibrium}
\label{fig:Displacement}
\end{figure*}

In the MRF T1-FF scan, the signal never reaches a steady state due to the varying sequence parameters, causing the contrast of the navigator images to vary throughout each repetition. This variability poses challenges for displacement calculation. To address this issue, the MRF navigator images were initially \textit{deseasonalized} \cite{boxTimeSeriesAnalysis2008} to eliminate the fluctuating MRF contrast. Deseasonalization involved decomposing the navigator temporal signals $I_{nav}(z,t)$ into three components: trend ($T$), seasonality ($S$), and residuals ($\epsilon$), as represented by the equation:
\begin{equation} \label{eq:Deseasonalization}
I_{nav}(z,t)=T(z,t)S(z,t)\epsilon(z,t)
\end{equation}
Here, $T$ represents the main signal trend over the entire scan duration, $S$ corresponds to the modulation of navigator images across each repetition of the MRF sequence, and $\epsilon$ denotes the residuals. $S$ is identical for each repetition of the MRF sequence, and was estimated for each position $z$ along the navigator readout by calculating a moving average to estimate the trend $T$ and then averaging the detrended signal across all repetitions of the MRF sequence.\\
\indent In the motion scan, displacement extraction allowed to bin the acquired data into M=5 respiratory phases with an equal number of spokes to maintain consistent undersampling levels across motion-resolved volumes (\cite{fengXDGRASPGoldenangleRadial2016}). Motion state 1 corresponded to full-expiration, while motion state 5 corresponded to full-inspiration. The edge z-coordinates of each motion state, denoted as $[e_{0},...,e_{M}]$, were used to sort the MRF scan data based on the displacement calculated using the deseasonalized MRF navigator images (Fig.\ref{fig:Displacement}).

\subsubsection{Deformation field estimation} \label{subsec:MoCoEstim}

The deformation field was estimated through an iterative process alternating between the reconstruction of magnitude volumes for all the respiratory phases of the motion scan, and the estimation of the deformation field from those volumes allowing to further refine the reconstruction. More specifically, volume reconstruction was formulated as a least-square minimization problem with soft-weighting (\cite{jaubertFreerunningCardiacMagnetic2020}), spatial Total Variation (TV) regularization, and cross-registration across all respiratory phases to enhance image quality (\cite{rank4DRespiratoryMotion2017}). 

This iterative process could be formalized by the following set of equations at step $t$:

\begin{equation} \label{eq:Motion}
M_{m\rightarrow m'}^{t}=argmin_{M} \mathcal{L}(MX_{m}^{t},X_{m'}^{t})+\mu Reg(M) 
\end{equation}

\begin{equation}\label{eq:Volumes}
X_{m}^{t+1} =argmin_{X}\sum_{m'} {||\sqrt{W_{m'}}(AM_{m\rightarrow m'}^{t}X-y)||_{2}^{2}} + \lambda ||TV(X)||_1
\end{equation}

Where $X_{m}^{t}$ is the reconstructed volume for motion state m at iteration t, $y$ is the acquired k-space data of the motion scan, $M_{m\rightarrow m'}^{t}$ is the motion field for registering respiratory phase m on respiratory phase m’ estimated at iteration t, $W_{m'}$ are the soft weights associated with respiratory phase m’ and A is the MRI acquisition encoding operator for the motion scan, combining radial k-space sampling, Fourier transform and coil sensitivities. The loss function $\mathcal{L}(.,.)$ penalizes differences between two images based on the normalized cross-correlation (\cite{avantsSymmetricDiffeomorphicImage2008}). This metric helped mitigate the impact of varying intensity across the image on the registration efficacy. The regularization cost $Reg(.)$ for ensuring a smooth deformation field was set as the L2-norm of the spatial gradient.

The soft weights $W_{m}$ were calculated using the following formula (\cite{jaubertFreerunningCardiacMagnetic2020}):

\begin{equation}\label{eq:SoftWeights}
W_{m}(n)=
    \begin{cases}
      1, & \text{if}\ d(n)\in [e_{m-1},e_{m}] \\
      e^{-\alpha(d(n)-e_{m})}, & \text{if} \ d(n)>e_{m} \text{ and} \ e^{-\alpha(d(n)-e_{m})}>\tau \\
      e^{-\alpha(e_{m-1}-d(n))}, & \text{if} \ d(n)<e_{m-1} \text{ and} \ e^{-\alpha(e_{m-1}-d(n))}>\tau \\
      0 ,& \text{otherwise}
    \end{cases}
\end{equation}
$\alpha$ was empirically set to 0.35 and $\tau$ to 0.5.

Equation \ref{eq:Volumes} was minimized via an in-house Python implementation of the conjugate gradient method, with number of iterations=5, $\lambda$=0.1. Equation 
\ref{eq:Motion} was solved using the Voxelmorph neural network as mentioned above (\cite{balakrishnanVoxelMorphLearningFramework2019}). 

Minimization problem \ref{eq:Motion} \ref{eq:Volumes} was initialized by separately reconstructing the different volumes without cross-registration. In each subsequent step, the reconstructed volumes for each motion state were fed into the UNet neural network VoxelMorph for estimating the deformation field (\cite{balakrishnanVoxelMorphLearningFramework2019}). The training set comprised all 2D slice pairs from adjacent respiratory phases, enabling the neural network to learn the incremental deformation field $M_{m\rightarrow m+1}$ between contiguous motion states $m$,$m+1$. These incremental deformation fields were then combined to calculate the deformation fields $M_{m\rightarrow m'}$ between all pairs of motion states $m$,$m'$. VoxelMorph was parameterized with 4 encoding and 6 decoding layers. The number of epochs was also empirically set to 1500 by observing that the loss changed by less than 0.1\% over 100 epochs after 1500 epochs of training. The regularization penalty $\mu$ was 0.05,
%Using a pre-trained network to reduce the registration time was as well tested.
while the number of iterations for the whole iterative process was 2. All reconstruction parameters were set empirically on representative examples. 

\subsubsection{Low-rank motion-corrected MRF reconstruction} \label{subsec:MoCoCorrection}
After estimation, the deformation field was applied iteratively to reconstruct the MRF complex singular volumes of the reference respiratory phase $m_0$. A Singular Value Decomposition (SVD) was performed on a dictionary of MRF simulated fingerprints. The MRF singular volumes were then generated by projecting the acquired data onto the 6 main components of this SVD (\cite{mcgivneySVDCompressionMagnetic2014}) explaining more than 99\% of the variance of the dictionary. The reconstruction process involved least-square minimization, integrating the motion information to ensure data consistency along with wavelet regularization (see Equation \ref{eq:SingularVolumes}). 
\begin{equation} \label{eq:SingularVolumes}
U_{m_{0}} =argmin_{U}\sum_{m'} {||\sqrt{\tilde{W_{m'}}}(\tilde{A}M_{m_{0}\rightarrow m'}U\Phi-\tilde{y})||_{2}^{2}} + \lambda_{W} ||W(U)||_1
\end{equation}

Where $U_{m_0}$ are the MRF singular volumes for respiratory phase $m_0$, $\tilde{y}$ is the acquired k-space data of the MRF T1-FF scan, $M_{m_{0}\rightarrow m'}$ is the final motion field for registering respiratory phase $m_0$ on respiratory phase $m'$ estimated on the motion scan, $\tilde{W_{m'}}$ are the weights associated with respiratory phase m’ and $\tilde{A}$ is the MRF acquisition encoding operator, $\Phi$ are the right singular vectors of the MRF temporal SVD, $W$ is the Daubechies wavelet analysis transform, with wavelet penalty $\lambda_{W}$=$1e^{-5}$ . The reference respiratory phase $m_0$ was chosen to correspond to the full expiration (resp. phase 1) as it is the most stable phase. 

Equation \ref{eq:SingularVolumes} was solved using an in-house Python implementation of the Fast Iterative Shrinkage-Thresholding Algorithm (FISTA) (\cite{beckFastIterativeShrinkageThresholding2009}). The number of iterations was empirically set to 3 by observing the convergence of the deformation field on representative examples. 

The motion-corrected singular volumes $U_{m_0}$ were then used for MRF pattern matching to reconstruct $T1_{H2O}$, FF, B0, B1 and $T1_{Fat}$  parametric maps for the reference respiratory phase $m_0$ using the bi-component dictionary matching method proposed in \cite{slioussarenkoBiComponentDictionary2024}.

\subsection{\textit{In vivo} MR acquisitions}

We validated the acquisition and reconstruction frameworks on an \textit{in vivo} dataset comprised of ten healthy volunteers (6 men, 4 women, mean age = 39 $\pm$ 12 years old) and one subject with Duchenne muscular dystrophy (DMD) (10 years old boy), one of the most common and severe forms of inherited muscular dystrophies \cite{salariGlobalPrevalenceDuchenne2022}. This protocol was approved by the local ethics committee (Comité de Protection des Personnes (CPP) Ile de France VI) and written informed consent was obtained. All MRI experiments were carried out on a 3T MRI ($Magnetom Prisma^{Fit}$, Siemens Healthineers, Erlangen, Germany). The subjects were lying in the head-first supine position. The body coil was used for RF transmission and signals were acquired using a 32-channel spine receive coil combined with a 18-channel flexible surface coil positioned on the volunteer's torso. The motion scan was acquired with the following sequence parameters: 1400 radial spokes per partition, TE/TR = 1.65/3.87 ms, nominal flip angle = 9º, bandwidth = 780 Hz/px, undersampling factor of 2 in the partition dimension, $T_{acq}$ = 5 min. The 3D MRF T1-FF sequence was acquired with varying TE, TR and nominal flip angles as described in \cite{martyQuantitativeSkeletalMuscle2021}, 1400 radial spokes/partition, bandwidth = 540 Hz/px, $T_{acq}$ = 10 min. Both acquisitions were performed in the sagittal view, with a field of view of 400 x 400 x 320 mm$^3$ and a spatial resolution of 1 x 1 x 5 mm$^3$. 

In the motion scan, sequence parameters TE and nominal flip angle were optimized through simulations to reduce scan time and maximize the signal difference between parenchyma and blood vessels in the liver, aiding in deformation field estimation (Fig.S1). The sharpness of the lung/liver interface on the navigator images was empirically optimized by varying TE and the selective pulse's slice thickness of the 1D gating spokes (Fig.S2).

%Regarding motion estimation from the navigators image, the displacements calculated on the motion scan and on the MRF scan were compared to show that they come from the same distribution, and hence that the motion fields calculated on the motion scan can be used on the MRF scan.%

\subsection{Validation metrics}
In evaluating the deformation field estimation, the registration quality was determined through visual comparisons between the registered volumes and the baseline. Additionally, Structural Similarity (SSIM) and Normalized Root Mean Squared Errors (NRMSE) were calculated between all respiratory phase and the reference full-expiration phase, both on the non-registered volumes and on the registered volumes. The convergence of the VoxelMorph neural network after 1500 epochs was assessed and compared against calculating the deformation fields on MRF singular volumes directly, as shown on one example (Fig. S3).
%The neural networks convergence was tested as a function of the number of epochs \ref{NNEpochsMapsComparison}. The generalization capacity of the neural network was assessed by applying pre-trained neural networks on other patients \ref{S3:PretrainedNNMapsComparison}.%

\begin{figure*}[!htbp]
\centering
\includegraphics[width=\textwidth]{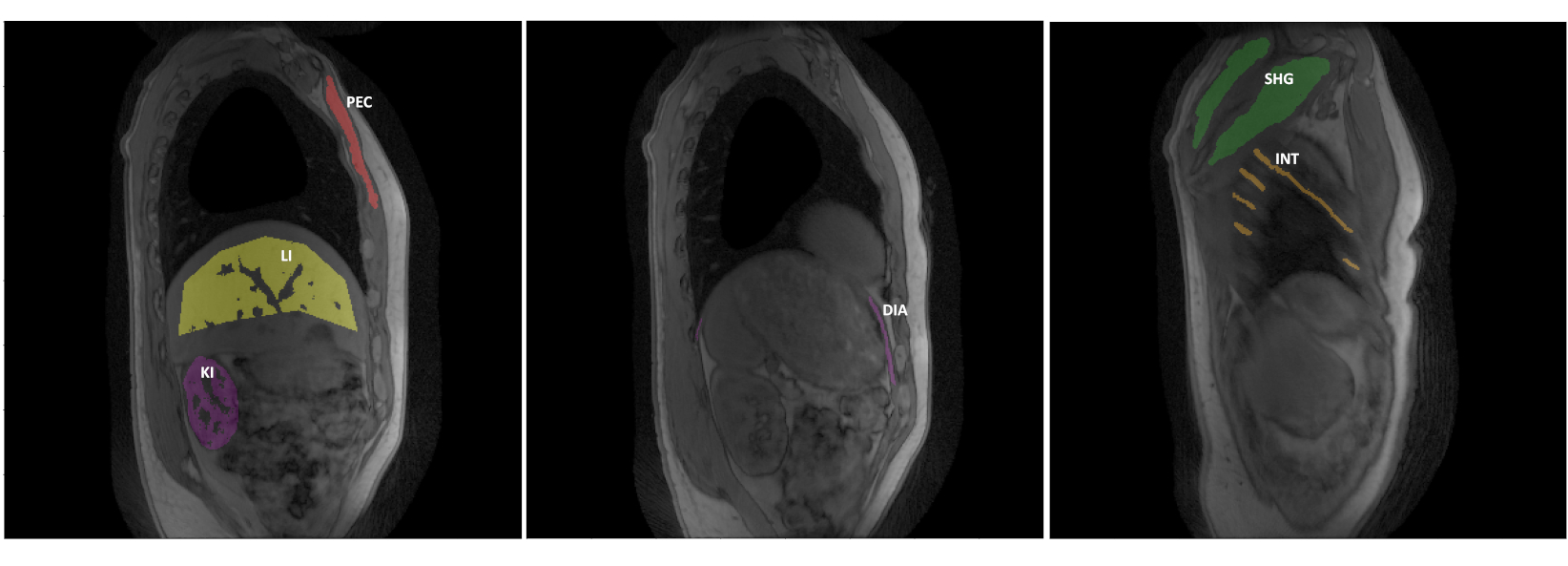}
\caption{Illustrative example of manual drawing of the different regions of interest for 3 sagittal slices of the motion scan in a healthy subject: Diaphragm (DIA), Intercostal muscles (INT), Kidney (KI), Liver (LI), \textit{Pectoralis major} (PEC), Shoulder girdle (SHG)}
\label{fig:ROI}
\end{figure*}

MoCo MRF T1-FF singular volumes and $T1_{H2O}$ and FF quantitative maps were initially assessed visually against 3D MRF T1-FF (Uncorrected MRF T1-FF), which reconstructs the Singular Volumes using a non-uniform Fourier Transform (nuFFT) \cite{barnettParallelNonuniformFast2018,barnettAliasingErrorExp2020} on all the acquired data without any binning or motion correction. B0, B1 and $T1_{Fat}$ parametric maps were also compared (Fig. S4).
Manual drawing of Regions of Interest (ROIs) was also performed in several muscles or muscle groups, namely the shoulder girdle (SHG), intercostal muscles (INT), \textit{Pectoralis major} (PEC), diaphragm (DIA), as well as in the liver (LI) and kidney (KI) (Fig.\ref{fig:ROI}). To quantitatively evaluate the accuracy and precision of the resulting quantitative maps, the distributions of $T1_{H2O}$ and FF values within each ROI were assessed and mean and standard deviations calculated. Since muscles like SHG, INT, and PEC are minimally affected by breathing motion, they were discernible in images obtained without MoCo. This allowed us to compare the ROI-level means and standard deviations of $T1_{H2O}$ and FF values with and without MoCo in these muscles to evaluate the potential bias on the resulting MoCo MRF T1-FF parametric maps. On the other hand, for muscles and organs significantly impacted by motion such as DIA, KI, and LI, ROIs could only be drawn consistently on MoCo MRF T1-FF images. %In these ROIs, we compared the mean and standard deviations of the $T1_{H2O}$ and FF distributions after MoCo with values found in the literature.

%We compared as well the end-to-end impact of some of the parameters that could allow to reduce the scan time in the future: impact of decreasing resolution on the deformation maps and interpolating it on the MRF scan resolution, impact of undersampling the MRF scan in partition (Figure S4).

\section{Results}
\label{Results}

\begin{figure*}[!htbp]
\centering
\includegraphics[width=\textwidth]{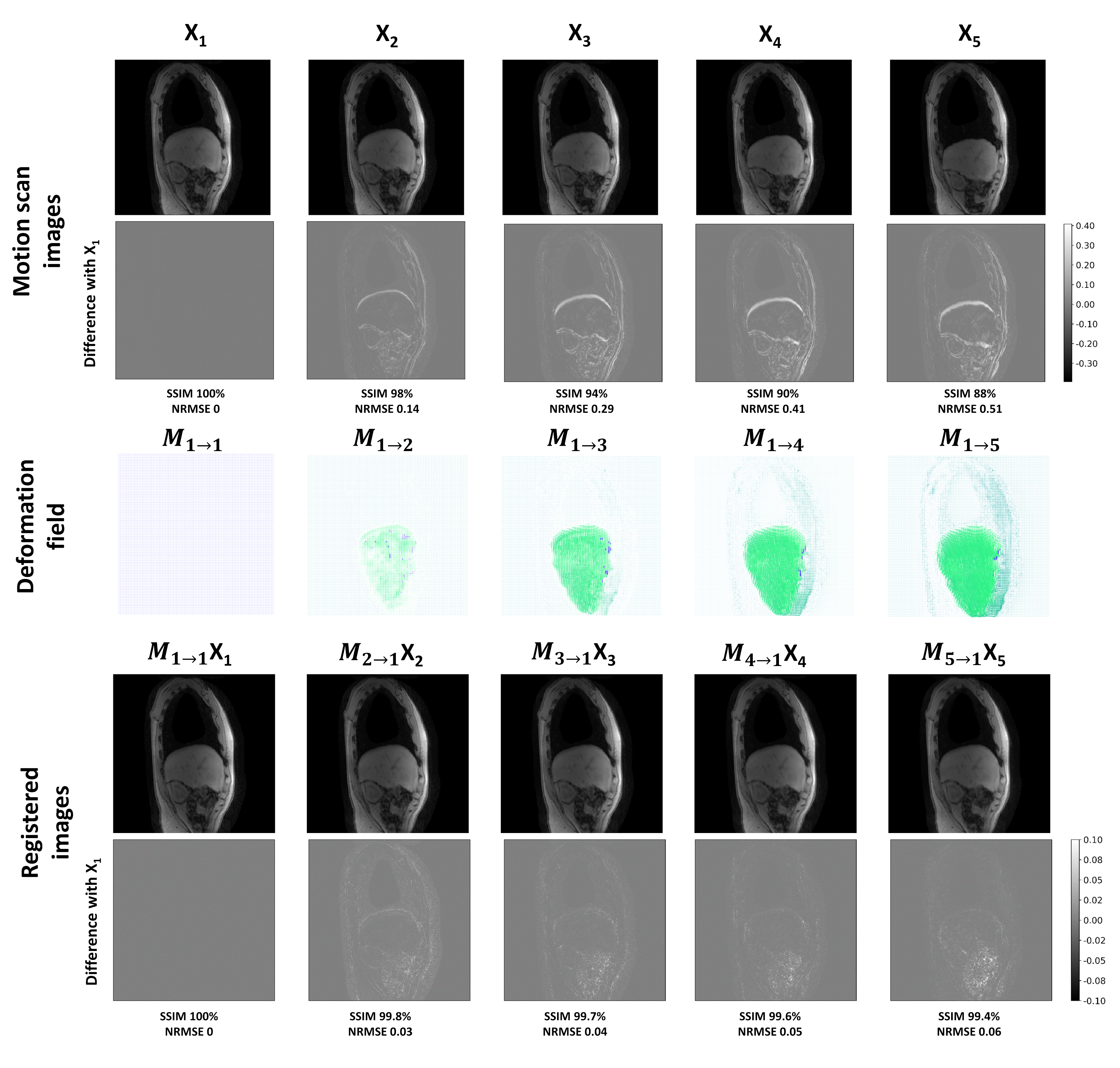}
\caption{Deformation field estimation and registration accuracy. The top two rows depict images reconstructed from the motion scan before registration, along with the difference compared to the reference full expiration respiratory phase (phase 1) for a single sagittal slice in a healthy subject. The third row displays the deformation fields estimated using VoxelMorph, while the last two rows illustrate the resulting difference images between the reference respiratory phase volume and the registered volume.}
\label{fig3:Registration}
\end{figure*}

The difference images between the reference respiratory phase and the other phases, with and without applying registration, show the effectiveness of the displacement field estimation on the motion scan \ref{eq:Motion} (Fig.\ref{fig3:Registration}). Supplementary videos (Fig.S5) demonstrating the registration results are also available for one healthy control. In the uncorrected images, noticeable movements included the liver descending from full expiration to full inspiration, and the abdominal ribcage shifting to the right. However, in the registered images, differences against the reference phase were greatly reduced for all motion states.
The mean and standard deviation of SSIM and NRMSE accross all subjects and slices were calculated and reported in table \ref{tab:SSIM}. The mean NRMSE of respiratory phases 2, 3, 4, and 5 against the reference respiratory phase 1 decreased from 0.10, 0.21, 0.30, and 0.36 to 0.04, 0.05, 0.07, and 0.07, respectively. Similarly, the mean SSIM index increased from 98.4\%, 95.8\%, 93.4\%, and 92.9\% to 99.6\%, 99.3\%, 99.1\%, and 99.3\%, respectively.

%The volumes rebuilt at iteration 0 and at last iteration for one example show the image improvement when using the iterative reconstruction process described in \ref{subsec:MoCoEstim}, reducing undersampling artefacts and improving contrast (Fig. S6). 

% Table generated by Excel2LaTeX from sheet 'Feuil1'
\begin{table}[!htbp]
  \centering
  \caption{Structural similarity (SSIM) and normalized root mean square error (NRMSE) between the volumes of respiratory phases 2-5 and reference respiratory phase (respiratory phase 1, corresponding to full expiration) for all subjects before and after motion registration. Data are presented as mean ± standard deviation.}
    \begin{tabular}{lll}
    \midrule
    \textbf{SSIM} & \textbf{Uncorrected motion scan volumes} & \textbf{Registered motion scan volumes} \\
    \midrule
    \textbf{Respiratory phase 2} & 98.4\% ± 2.2\% & 99.6\% ± 0.7\% \\
    \textbf{Respiratory phase 3} & 95.8\% ± 3.0\% & 99.3\% ± 1.0\% \\
    \textbf{Respiratory phase 4} & 93.4\% ± 3.4\% & 99.1\% ± 1.4\% \\
    \textbf{Respiratory phase 5} & 92.9\% ± 2.6\% & 99.3\% ± 0.7\% \\
    \midrule
    \textbf{NRMSE} & \textbf{Uncorrected motion scan volumes} & \textbf{Registered motion scan volumes} \\
    \midrule
    \textbf{Respiratory phase 2} & 0.10 ± 0.04 & 0.04 ± 0.02 \\
    \textbf{Respiratory phase 3} & 0.21 ± 0.07 & 0.05 ± 0.02 \\
    \textbf{Respiratory phase 4} & 0.30 ± 0.10 & 0.07 ± 0.02 \\
    \textbf{Respiratory phase 5} & 0.36 ± 0.11 & 0.07 ± 0.02 \\
    
    \end{tabular}%
  \label{tab:SSIM}%
\end{table}%

\begin{figure*}[!htbp]
\centering
\includegraphics[width=\textwidth]{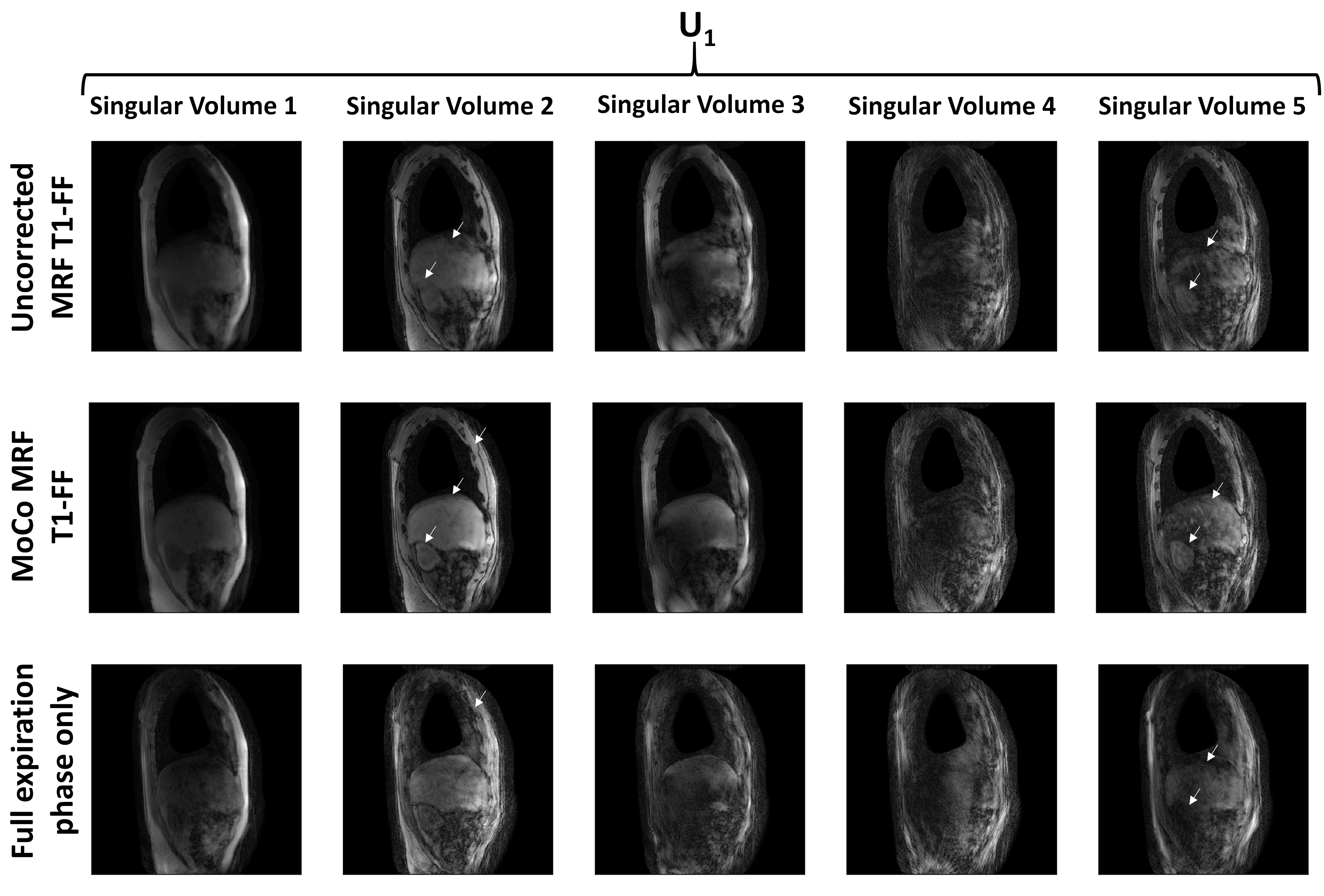}
\caption{The first five singular volumes reconstructed from the uncorrected MRF T1-FF sequence reconstructing the singular volumes from a nuFFT on all acquired data without binning or motion correction, the motion-corrected (MoCo) MRF T1-FF data, and from the MRF T1-FF data acquired during the reference respiratory phase only (corresponding to the full-expiration phase). White arrows highlight areas where motion blurring, undersampling, and weak contrast are notably reduced following motion correction.}
\label{fig4:Singular}
\end{figure*}

Applying the deformation fields calculated from the motion scan to reconstruct the MRF singular volumes using Equation \ref{eq:SingularVolumes} resulted in sharper singular volumes compared to the uncorrected method (Fig.\ref{fig4:Singular}). In this representative subject, singular volumes reconstructed solely from the full-expiration respiratory phase exhibited pronounced undersampling artifacts and weaker contrast. This impact was particularly evident in singular volumes 2 (representing phases where fat and water are out of phase) and 5 (providing high T1 contrast), as indicated by the white arrows. In the uncorrected images, motion-induced blurring in the kidney and liver regions affected the clarity of the 2$^{nd}$ singular volume, while utilizing only the full-expiration data partially mitigated this blurring but introduced streaking artifacts, notably in the pectoral muscles. Conversely, in the 5$^{th}$ singular volume, exclusive use of the full-expiration data led to the disappearance of T1 contrast between vessels and liver, whereas these structures remained clearly visible in the MoCo MRF T1-FF images.

\begin{figure*}[!htbp]
\centering
\includegraphics[width=\textwidth]{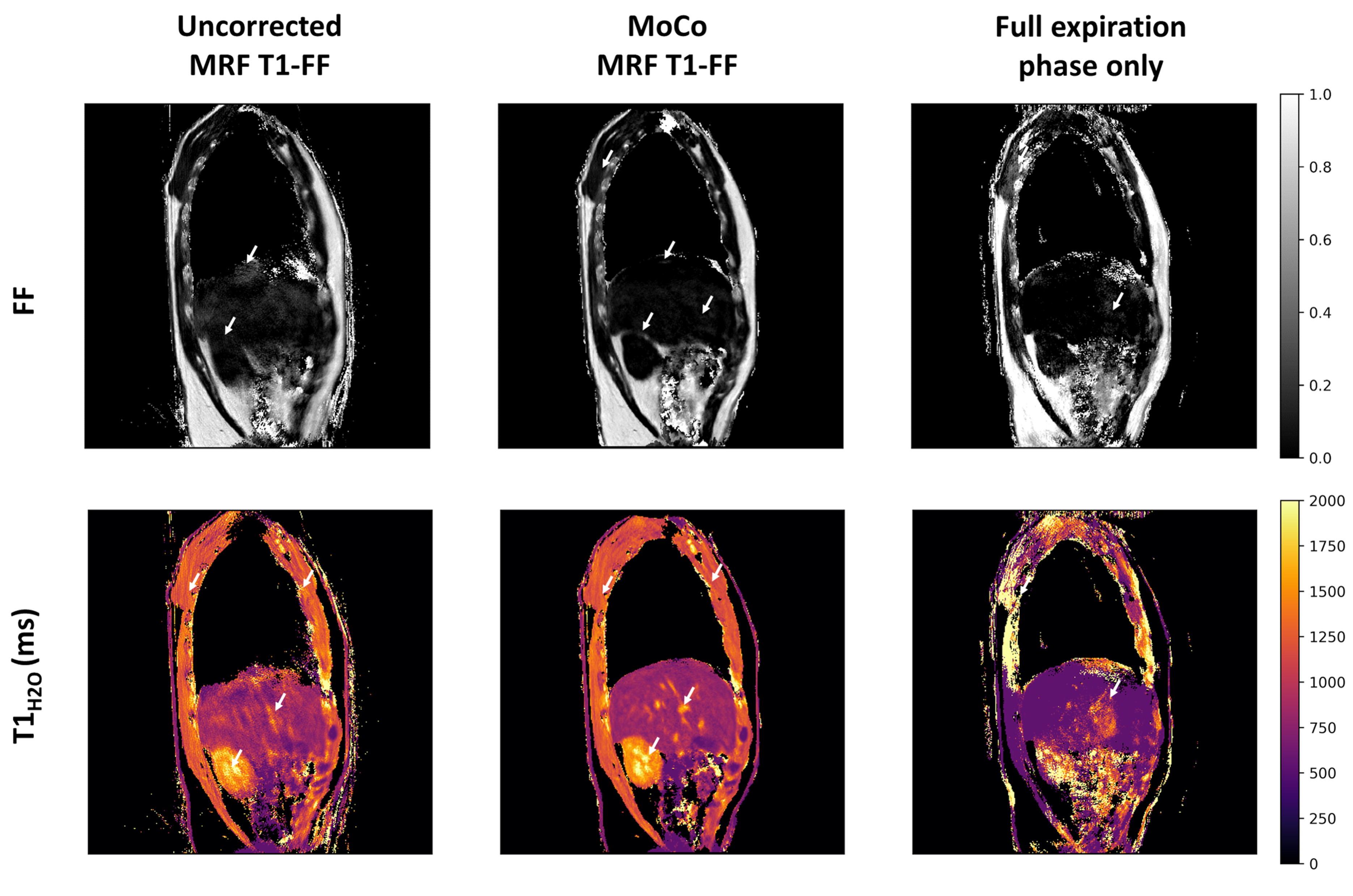}
\caption{Representative parametric FF and $T1_{H2O}$ maps obtained at one slice level in a healthy volunteer derived from the uncorrected and motion-corrected (MoCo) MRF T1-FF reconstructions, and from the MRF T1-FF data acquired during the reference respiratory phase only (corresponding to the full-expiration phase). White arrows emphasize regions where motion blurring, parameter estimation bias and streaking artifacts are significantly mitigated with the motion correction method.}
\label{fig5:Liver}
\end{figure*}

\begin{figure*}[!htbp]
\centering
\includegraphics[width=\textwidth,,height=\textheight,keepaspectratio]{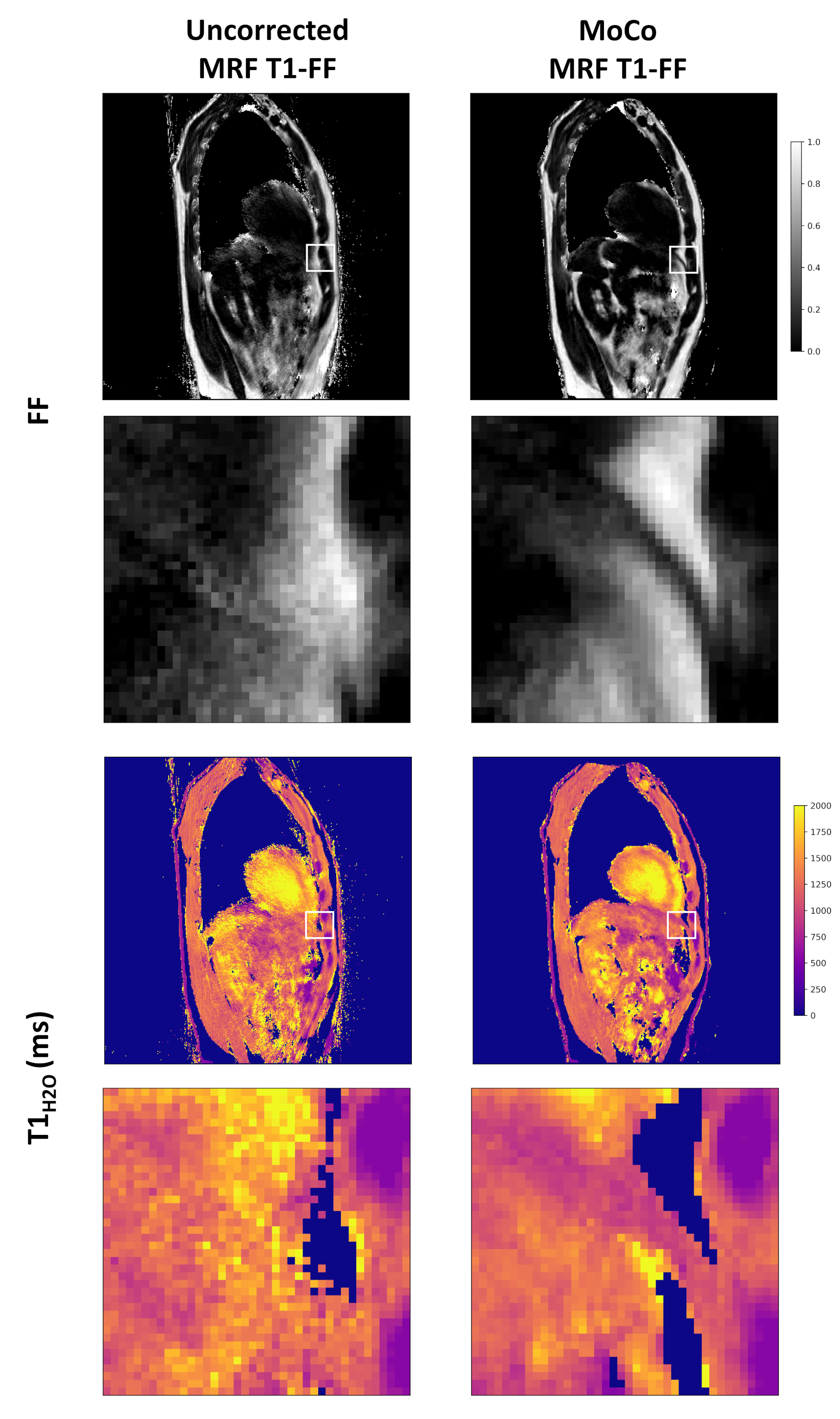}
\caption{Representative parametric FF and $T1_{H2O}$ maps obtained at one slice level in a healthy volunteer derived from both the uncorrected and motion-corrected (MoCo) MRF T1-FF reconstructions. The second and fourth rows present zoomed images corresponding to the white squares on the full images. Following motion correction, a thin layer of muscle is discernible on the FF and $T1_{H2O}$ maps, corresponding to the diaphragm.}
\label{fig6:Diaphragm}
\end{figure*}

As a result of this motion correction, the MRF parametric maps exhibited significantly enhanced sharpness and reduced artefacts (Fig.\ref{fig5:Liver}). Notably, on the $T1_{H2O}$ maps, distinct anatomical features of the liver such as the dome and vessels become more discernible, while FF decreased and exhibited greater uniformity. Moreover, the kidney calyx was clearly delineated on the $T1_{H2O}$ maps, and motion-induced blurring in the renal medulla vanished on the FF maps. Streaking artifacts in the pectoral, dorsal, and abdominal muscles were noticeably diminished. Reconstruction using only spokes from the full expiration phase showed severe undersampling artefacts and strong bias in FF and $T1_{H2O}$ estimation. Furthermore, the diaphragm could be detected on several slices of the MoCo MRF T1-FF maps, indicated by a thin arch devoid of fat between the liver and the heart (Fig.\ref{fig6:Diaphragm}). This region corresponded to a slender muscle layer clearly outlined on the $T1_{H2O}$ maps.

\begin{figure*}[!htbp]
\centering
\includegraphics[width=\textwidth,,height=\textheight,keepaspectratio]{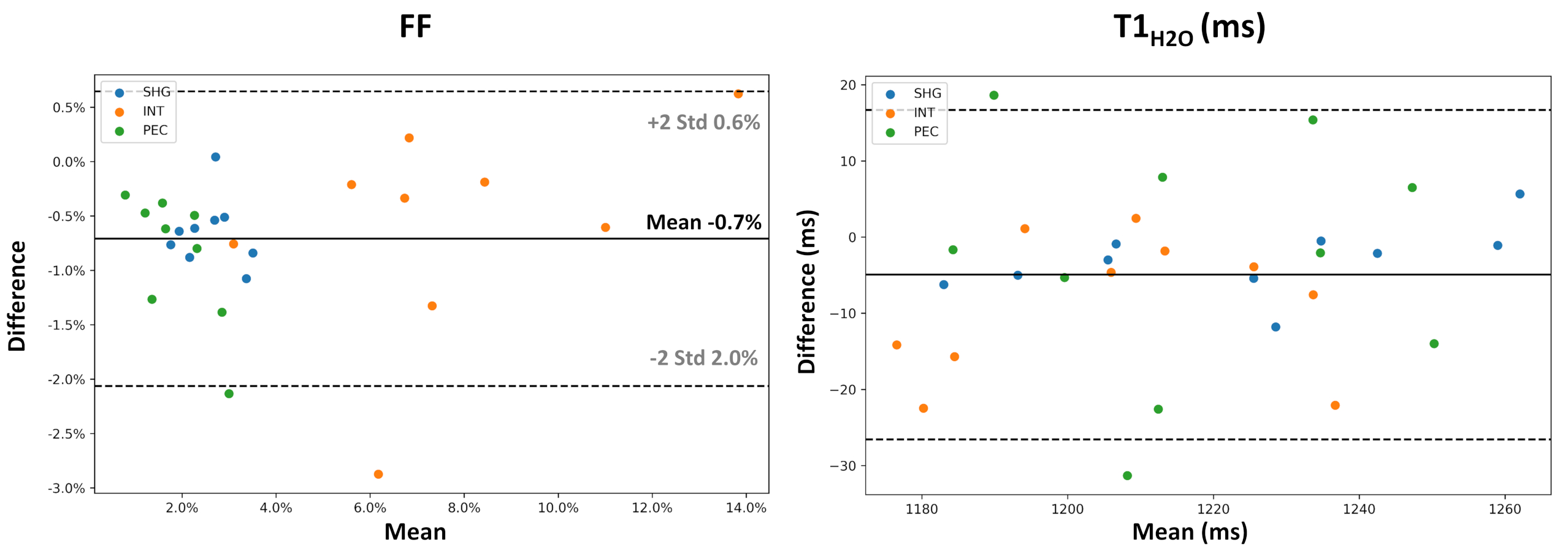}
\caption{Bland-Altman plot comparing the ROI-level fat fraction (FF) and water T1 ($T1_{H2O}$) values derived from both the uncorrected and motion-corrected (MoCo) MRF T1-FF reconstruction methods in the shoulder girdle (SHG), intercostal muscles (INT), and \textit{Pectoralis major} (PEC). Each data point represents an individual subject. The solid line represents the mean difference between the values obtained with both reconstructions, while the dashed lines indicate the 95\% confidence interval.}
\label{fig7:BlandAltman}
\end{figure*}

\begin{figure*}[!htbp]
\centering
\includegraphics[width=\textwidth,,height=\textheight,keepaspectratio]{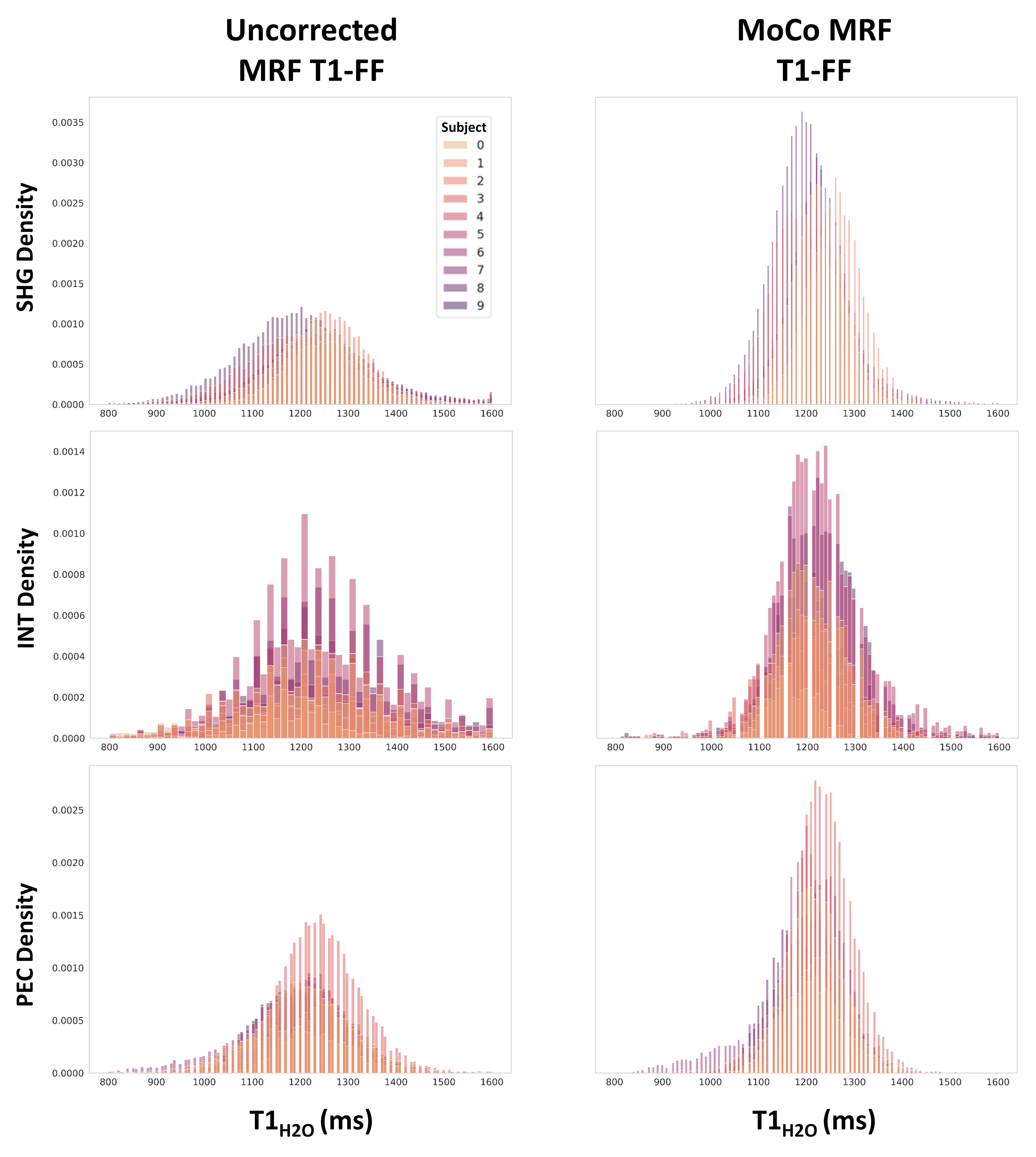}
\caption{Distribution of fat fraction (FF) and water T1 ($T1_{H2O}$) values obtained in the shoulder girdle (SHG), intercostal muscles (INT), and \textit{Pectoralis major} (PEC) from the uncorrected and the motion-corrected (MoCo) MRF T1-FF reconstruction. Each color represents an individual subject.}
\label{fig8:Precision}
\end{figure*}

The distribution of FF and $T1_{H2O}$ values were also assessed at the ROI level. For SHG, INT, and PEC, which are only minimally affected by motion, the Bland-Altman plot indicated no significant bias between Uncorrected MRF T1-FF and MoCo MRF T1-FF for these muscles, with a mean difference of -0.7\% for FF and -4.9ms for $T1_{H2O}$ values (Fig.\ref{fig7:BlandAltman}). In terms of precision, we analyzed the distribution of $T1_{H2O}$ for all subjects in these muscles (Fig.\ref{fig8:Precision}). MoCo MRF T1-FF significantly improved precision, as evidenced by the narrower distributions across all healthy controls. 
%Grouping SHG, INT and PEC and all healthy controls, the mean $T1_{H2O}$ went from 1223 to 1217 ms with MoCo MRF T1-FF, with the standard deviation dropping from 111 to 70ms. FF precision improved as well, with a standard deviation going down from 4.8\% to 3.8\% with MoCo MRF T1-FF. Mean FF slightly decreased from 3.4\% to 2.6\%.

Table \ref{tab:T1ROIResultsControlsMuscle} presents the ROI-level mean and standard deviation for SHG, INT, and PEC across all subjects. The mean values were consistent between both methods, although FF values were slightly higher with the uncorrected method. Standard deviations for both FF and $T1_{H2O}$ decreased across all ROIs and subjects with the MoCo MRF T1-FF reconstruction. However, the degree of precision enhancement varied depending on the subjects' breathing characteristics and the specific ROI. FF and $T1_{H2O}$ values showed alignment across healthy controls (Subjects 0-9) for SHG and PEC, while slight heterogeneity was observed for INT, with FF reaching up to 14\% for Subject 8. In contrast, the subject with DMD exhibited high FF in the SHG and PEC muscles (39\% and 34\% respectively) but no fat replacement in the intercostal muscles. Mean $T1_{H2O}$ values also appeared slightly elevated and with higher standard deviation for the subject with DMD compared to healthy controls.

For targets significantly affected by breathing motion, we present the mean and standard deviation at the ROI level using the MoCo MRF T1-FF method in Table \ref{tab:T1ROIResultsControlsOther}. In the diaphragm, FF values were slightly higher than those observed in stationary muscles, while mean $T1_{H2O}$ values aligned with those measured in SHG and PEC muscles, albeit with a seemingly higher standard deviation. FF values for the liver and kidneys remained consistently low across all subjects. $T1_{H2O}$ values were consistently higher in the kidneys than in the muscles and liver.

\begin{table*}[!htbp]
  \centering
  \caption{ROI-level fat fraction (FF) and water T1 ($T1_{H2O}$) values (mean ± standard deviation)  measured in the 10 healthy volunteers (Subjects 0-9) and the subject with Duchenne muscular dystrophy (Subject 10) for the uncorrected and motion-corrected (MoCo) MRF T1-FF reconstructions assessed in the shoulder girdle (SHG), intercostal muscles (INT), and \textit{Pectoralis major} (PEC) muscles. Illustrative literature values were added in the last row of each table section.}
  \begin{adjustbox}{width=1\textwidth}
  \begin{threeparttable}
    \begin{tabular}{rlcccccc}
          &       & \multicolumn{1}{l}{\textbf{SHG}} &       & \multicolumn{1}{l}{\textbf{INT}} &       & \multicolumn{1}{l}{\textbf{PEC}} &  \\
          &       & \multicolumn{1}{l}{Uncorrected MRF T1-FF} & \multicolumn{1}{l}{MoCo MRF T1-FF} & \multicolumn{1}{l}{Uncorrected MRF T1-FF} & \multicolumn{1}{l}{MoCo MRF T1-FF} & \multicolumn{1}{l}{Uncorrected MRF T1-FF} & \multicolumn{1}{l}{MoCo MRF T1-FF} \\
\cmidrule{1-8}    \multicolumn{1}{l}{\textbf{$T1_{H2O}$ (ms)}} & Subject 0 & \multicolumn{1}{l}{1236 ± 96} & \multicolumn{1}{l}{1222 ± 55} & \multicolumn{1}{l}{1208 ± 146} & \multicolumn{1}{l}{1210 ± 71} & \multicolumn{1}{l}{1210 ± 101} & \multicolumn{1}{l}{1216 ± 59} \\
          & Subject 1 & \multicolumn{1}{l}{1244 ± 82} & \multicolumn{1}{l}{1241 ± 55} & \multicolumn{1}{l}{1184 ± 83} & \multicolumn{1}{l}{1169 ± 49} & \multicolumn{1}{l}{1181 ± 118} & \multicolumn{1}{l}{1199 ± 81} \\
          & Subject 2 & \multicolumn{1}{l}{1260 ± 81} & \multicolumn{1}{l}{1264 ± 58} & \multicolumn{1}{l}{1251 ± 123} & \multicolumn{1}{l}{1224 ± 80} & \multicolumn{1}{l}{1258 ± 77} & \multicolumn{1}{l}{1243 ± 61} \\
          & Subject 3 & \multicolumn{1}{l}{1186 ± 116} & \multicolumn{1}{l}{1179 ± 65} & \multicolumn{1}{l}{1197 ± 148} & \multicolumn{1}{l}{1175 ± 84} & \multicolumn{1}{l}{1236 ± 92} & \multicolumn{1}{l}{1233 ± 65} \\
          & Subject 4 & \multicolumn{1}{l}{1260 ± 82} & \multicolumn{1}{l}{1258 ± 52} & \multicolumn{1}{l}{1216 ± 136} & \multicolumn{1}{l}{1211 ± 80} & \multicolumn{1}{l}{1245 ± 91} & \multicolumn{1}{l}{1250 ± 50} \\
          & Subject 5 & \multicolumn{1}{l}{1236 ± 134} & \multicolumn{1}{l}{1224 ± 77} & \multicolumn{1}{l}{1238 ± 142} & \multicolumn{1}{l}{1226 ± 92} & \multicolumn{1}{l}{1224 ± 78} & \multicolumn{1}{l}{1201 ± 43} \\
          & Subject 6 & \multicolumn{1}{l}{1208 ± 106} & \multicolumn{1}{l}{1204 ± 69} & \multicolumn{1}{l}{1194 ± 128} & \multicolumn{1}{l}{1168 ± 78} & \multicolumn{1}{l}{1180 ± 113} & \multicolumn{1}{l}{1180 ± 94} \\
          & Subject 7 & \multicolumn{1}{l}{1241 ± 127} & \multicolumn{1}{l}{1235 ± 67} & \multicolumn{1}{l}{1214 ± 131} & \multicolumn{1}{l}{1204 ± 74} & \multicolumn{1}{l}{1226 ± 75} & \multicolumn{1}{l}{1241 ± 44} \\
          & Subject 8 & \multicolumn{1}{l}{1198 ± 128} & \multicolumn{1}{l}{1190 ± 67} & \multicolumn{1}{l}{1248 ± 142} & \multicolumn{1}{l}{1230 ± 84} & \multicolumn{1}{l}{1202 ± 95} & \multicolumn{1}{l}{1196 ± 70} \\
          & Subject 9 & \multicolumn{1}{l}{1208 ± 95} & \multicolumn{1}{l}{1206 ± 53} & \multicolumn{1}{l}{1196 ± 122} & \multicolumn{1}{l}{1196 ± 86} & \multicolumn{1}{l}{1225 ± 96} & \multicolumn{1}{l}{1192 ± 63} \\
          & Subject 10 (DMD) & \multicolumn{1}{l}{1278 ± 166} & \multicolumn{1}{l}{1274 ± 128} & \multicolumn{1}{l}{1295 ± 167} & \multicolumn{1}{l}{1264 ± 110} & \multicolumn{1}{l}{1211 ± 218} & \multicolumn{1}{l}{1225 ± 186} \\
\cmidrule{2-8}          & Literature & \multicolumn{6}{c}{1198 ± 50 \cite{martyWaterfatSeparationMR2021}} \\

\cmidrule{1-8}    \multicolumn{1}{l}{\textbf{FF}} & Subject 0 & \multicolumn{1}{l}{0.03 ± 0.03} & \multicolumn{1}{l}{0.02 ± 0.02} & \multicolumn{1}{l}{0.08 ± 0.09} & \multicolumn{1}{l}{0.05 ± 0.05} & \multicolumn{1}{l}{0.04 ± 0.05} & \multicolumn{1}{l}{0.02 ± 0.03} \\
          & Subject 1 & \multicolumn{1}{l}{0.02 ± 0.03} & \multicolumn{1}{l}{0.02 ± 0.02} & \multicolumn{1}{l}{0.03 ± 0.03} & \multicolumn{1}{l}{0.03 ± 0.03} & \multicolumn{1}{l}{0.04 ± 0.05} & \multicolumn{1}{l}{0.02 ± 0.03} \\
          & Subject 2 & \multicolumn{1}{l}{0.03 ± 0.03} & \multicolumn{1}{l}{0.03 ± 0.03} & \multicolumn{1}{l}{0.09 ± 0.07} & \multicolumn{1}{l}{0.08 ± 0.07} & \multicolumn{1}{l}{0.01 ± 0.02} & \multicolumn{1}{l}{0.01 ± 0.02} \\
          & Subject 3 & \multicolumn{1}{l}{0.04 ± 0.04} & \multicolumn{1}{l}{0.03 ± 0.03} & \multicolumn{1}{l}{0.08 ± 0.07} & \multicolumn{1}{l}{0.07 ± 0.06} & \multicolumn{1}{l}{0.03 ± 0.03} & \multicolumn{1}{l}{0.02 ± 0.03} \\
          & Subject 4 & \multicolumn{1}{l}{0.03 ± 0.04} & \multicolumn{1}{l}{0.02 ± 0.03} & \multicolumn{1}{l}{0.07 ± 0.07} & \multicolumn{1}{l}{0.07 ± 0.06} & \multicolumn{1}{l}{0.01 ± 0.02} & \multicolumn{1}{l}{0.01 ± 0.02} \\
          & Subject 5 & \multicolumn{1}{l}{0.02 ± 0.04} & \multicolumn{1}{l}{0.01 ± 0.03} & \multicolumn{1}{l}{0.11 ± 0.10} & \multicolumn{1}{l}{0.11 ± 0.09} & \multicolumn{1}{l}{0.02 ± 0.03} & \multicolumn{1}{l}{0.01 ± 0.02} \\
          & Subject 6 & \multicolumn{1}{l}{0.03 ± 0.04} & \multicolumn{1}{l}{0.03 ± 0.03} & \multicolumn{1}{l}{0.06 ± 0.06} & \multicolumn{1}{l}{0.05 ± 0.04} & \multicolumn{1}{l}{0.02 ± 0.03} & \multicolumn{1}{l}{0.01 ± 0.02} \\
          & Subject 7 & \multicolumn{1}{l}{0.03 ± 0.05} & \multicolumn{1}{l}{0.02 ± 0.03} & \multicolumn{1}{l}{0.07 ± 0.06} & \multicolumn{1}{l}{0.07 ± 0.05} & \multicolumn{1}{l}{0.02 ± 0.03} & \multicolumn{1}{l}{0.01 ± 0.02} \\
          & Subject 8 & \multicolumn{1}{l}{0.04 ± 0.04} & \multicolumn{1}{l}{0.03 ± 0.03} & \multicolumn{1}{l}{0.14 ± 0.11} & \multicolumn{1}{l}{0.14 ± 0.09} & \multicolumn{1}{l}{0.03 ± 0.04} & \multicolumn{1}{l}{0.02 ± 0.03} \\
          & Subject 9 & \multicolumn{1}{l}{0.05 ± 0.05} & \multicolumn{1}{l}{0.04 ± 0.04} & \multicolumn{1}{l}{0.08 ± 0.09} & \multicolumn{1}{l}{0.07 ± 0.05} & \multicolumn{1}{l}{0.05 ± 0.05} & \multicolumn{1}{l}{0.04 ± 0.05} \\
          & Subject 10 (DMD) & \multicolumn{1}{l}{0.39 ± 0.27} & \multicolumn{1}{l}{0.39 ± 0.27} & \multicolumn{1}{l}{0.06 ± 0.08} & \multicolumn{1}{l}{0.05 ± 0.06} & \multicolumn{1}{l}{0.33 ± 0.19} & \multicolumn{1}{l}{0.34 ± 0.15} \\
\cmidrule{2-8}          & Literature & \multicolumn{6}{c}{Between 0.034 ± 0.010 and 0.068 ± 0.019 \cite{lopezkolkovskyMultiparametricAgingStudy2024}} \\
    \end{tabular}%
    \end{threeparttable}
    \end{adjustbox}
  \label{tab:T1ROIResultsControlsMuscle}%
\end{table*}%

\begin{table*}[!htbp]
  \centering
  \caption{ROI-level fat fraction (FF) and water T1 ($T1_{H2O}$) values (mean ± standard deviation) measured in the 10 healthy volunteers (Subjects 0-9) and the subject with Duchenne muscular dystrophy (Subject 10) for the uncorrected and motion-corrected (MoCo) MRF T1-FF reconstructions assessed in the diaphragm (DIA), kidneys (KI) and liver (LI). Illustrative literature values were added in the last row of each table section.}
  \begin{adjustbox}{width=1\textwidth}
  \begin{threeparttable}
    \begin{tabular}{rllll}
          &       & \textbf{DIA} & \textbf{LI} & \textbf{KI} \\
          &       & MoCo MRF T1-FF  & MoCo MRF T1-FF  & MoCo MRF T1-FF \\
\cmidrule{1-5}    \multicolumn{1}{l}{\textbf{$T1_{H2O}$ (ms)}} & Subject 0 & 1196 ± 100 & 868 ± 144 & 1446 ± 175 \\
          & Subject 1 & 1186 ± 64 & 958 ± 135 & 1482 ± 88 \\
          & Subject 2 & 1325 ± 159 & 927 ± 135 & 1478 ± 242 \\
          & Subject 3 & 1225 ± 121 & 871 ± 139 & 1412 ± 183 \\
          & Subject 4 & 1242 ± 102 & 978 ± 116 & 1460 ± 238 \\
          & Subject 5 & 1227 ± 117 & 821 ± 161 & 1465 ± 206 \\
          & Subject 6 & 1231 ± 109 & 808 ± 136 & 1419 ± 178 \\
          & Subject 7 & 1155 ± 134 & 899 ± 147 & 1395 ± 218 \\
          & Subject 8 & 1206 ± 103 & 865 ± 133 & 1492 ± 150 \\
          & Subject 9 & 1265 ± 127 & 1017 ± 125 & 1475 ± 109 \\
          & Subject 10 (DMD) & 1164 ± 131 & 892 ± 119 & 1330 ± 171 \\
\cmidrule{2-5}          & Literature & 1198 ± 50 \cite{martyWaterfatSeparationMR2021} & 873 ± 96 \cite{fellnerWaterFatSeparatedT12023} & 1314 ± 77 \cite{chenMRFingerprintingRapid2016} \\
\cmidrule{1-5}    \multicolumn{1}{l}{\textbf{FF}} & Subject 0 & 0.10 ± 0.10 & 0.03 ± 0.04 & 0.07 ± 0.14 \\
          & Subject 1 & 0.01 ± 0.03 & 0.01 ± 0.02 & 0.03 ± 0.06 \\
          & Subject 2 & 0.08 ± 0.08 & 0.01 ± 0.02 & 0.06 ± 0.11 \\
          & Subject 3 & 0.08 ± 0.10 & 0.05 ± 0.04 & 0.06 ± 0.12 \\
          & Subject 4 & 0.10 ± 0.07 & 0.08 ± 0.04 & 0.08 ± 0.14 \\
          & Subject 5 & 0.12 ± 0.11 & 0.03 ± 0.04 & 0.10 ± 0.14 \\
          & Subject 6 & 0.09 ± 0.09 & 0.04 ± 0.03 & 0.07 ± 0.12 \\
          & Subject 7 & 0.08 ± 0.08 & 0.01 ± 0.03 & 0.08 ± 0.15 \\
          & Subject 8 & 0.08 ± 0.09 & 0.01 ± 0.02 & 0.03 ± 0.07 \\
          & Subject 9 & 0.11 ± 0.11 & 0.01 ± 0.02 & 0.04 ± 0.07 \\
          & Subject 10 (DMD) & 0.06 ± 0.08 & 0.02 ± 0.03 & 0.04 ± 0.08 \\
\cmidrule{2-5}          & Literature & Between 0.034 ± 0.010 and 0.068 ± 0.019 \cite{lopezkolkovskyMultiparametricAgingStudy2024}   & [0.01-0.08] \cite{shinNormalRangeHepatic2015} & N/A \\
    \end{tabular}%
    \end{threeparttable}
    \end{adjustbox}
  \label{tab:T1ROIResultsControlsOther}%
\end{table*}%

%\subsection{Motion calculation}
%\begin{itemize}
%\item \textbf{Displacement calculation illustration - shifted images plot ?}
%\item \textbf{Comparison of displacement on MRF and motion scan - show they come from same distribution }
%\item Parameters of motion calculation ?
%\end{itemize}

%\subsection{Acquisition parameters optimization}
%\begin{itemize}
%\item Contrast of vessel and liver as a function of TE and theta (illustrative liver image)
%\item Gating spokes contrast as a function of TE and slice thickness
%\end{itemize}

%\subsection{Motion field estimation}
%\begin{itemize}
%\item Neural network parameters optimization 
%\item \textbf{Motion field illustration and movie of registered vs non-registered images}
%\item \textbf{Comparison of results between trained and pre-trained neural network}
%\end{itemize}

%\subsection{Quantitative maps results}
%\begin{itemize}
%\item \textbf{Comparison of maps no correction vs correction - boxplot + std ..}
%\item \textbf{impact of deformation map downsampling on quantitative maps - boxplot per ROI and visual maps}%
%\item \textbf{impact of MRF undersampling - boxplot per ROI and visual maps}
%\item impact of binning - boxplot per ROI and visual maps
%\end{itemize}

\section{Discussion}
\label{Discussion}

This study introduces a dedicated MRI acquisition and reconstruction framework tailored for robust and motion-corrected reconstruction of high-resolution 3D multi-parametric maps of the upper-body acquired under free-breathing conditions. This was successfully applied to quantify $T1_{H2O}$ and FF in this anatomical region on a small group of ten healthy volunteers and one subject with a neuromuscular disease. Our approach effectively mitigated blurring caused by respiratory motion as well as undersampling artifacts, resulting in high-quality 3D parametric maps.

Our pipeline involved dividing the acquisition process into two stages: a preliminary motion scan followed by a 3D magnetic resonance fingerprinting sequence. In the context of quantitative 3D cardiac and liver imaging, \cite{cruzGeneralizedLowRank2022} proposed a generalized low-rank approach to simultaneously reconstruct the deformation fields and the parametric maps from a single 3D MRF sequence. This approach allowed for the reconstruction of auxiliary motion-resolved volumes, which were subsequently co-registered using non-rigid motion deformation. By combining these deformation fields with low-rank reconstruction, motion-corrected T1 and T2 MRF maps were generated. However, this method required the selection of specific segments of the MRF time-series which display optimal contrast in the reconstructed auxiliary images for efficient registration of the respiratory phases. In contrast, our modular approach offers a significant advantage by enabling separate optimization of the motion and MRF scans to fulfill their specific purposes – motion field estimation for the motion scan and parametric maps reconstruction for MRF. For instance, to achieve precise estimation of deformation fields in the liver, we used specific flip angles and echo times (TE) to enhance T1 contrast between the liver and the vessels (\cite{ernstApplicationFourierTransform1966}). Meanwhile, the acquisition scheme of the MRF T1-FF scan remained mostly unaltered compared to the proposed version specifically tailored for $T1_{H2O}$ and FF parametric mapping (\cite{martyQuantitativeSkeletalMuscle2021}).

The only modification made to the MRF sequence involved the integration of 1D navigators at intervals of every 28 spokes to detect respiratory motion. This method was chosen over alternatives such as respiratory bellows or the Pilot Tone device due to its simplicity, as it eliminates the need for additional external hardware and ensures inherent synchronization with the MRI acquisition. The main challenge in combining gating spokes with MRF lied in accurately distinguishing between signal variations induced by the MRF pattern and those caused by motion in the navigator time series. To address this challenge, we employed a statistical approach to eliminate the 'seasonal MRF component' from the navigator time series (\cite{boxTimeSeriesAnalysis2008}). This method, commonly utilized in economics to isolate the trend of a data time series from cyclical seasonal variations, proved also effective in our context. While untested, the Pilot Tone device could potentially replace gating spokes in both the motion scan and the MRF scan to detect respiratory motion (\cite{solomonFreeBreathingRadial2021}). This substitution would offer the advantage of enhanced modularity, as it eliminates the need for modifications to both acquisition sequences. However, this comes at the expense of requiring additional hardware.

%Regarding the image reconstruction step, our approach bins the acquired data into different respiratory phases and applies motion deformation fields to register the MRF singular volumes onto a reference motion state. This correction process minimizes a least-square data consistency problem by incorporating the non-rigid deformation information. In our framework, deformation fields were estimated using VoxelMorph, an open-source UNet neural network for image registration (\cite{balakrishnanVoxelMorphLearningFramework2019}) on the motion scan data. 
Our study demonstrated the effectiveness of VoxelMorph in accurately estimating deformation fields between motion-resolved images acquired in different respiratory phases in our small group of volunteers, yielding to excellent NRMSE and SSIM indices between the different co-registered volumes consistently in all subjects. Employing a neural network offers two potential advantages for future applications. Firstly, pre-training on an image database could facilitate rapid registration, thus speeding up the reconstruction which lasted several hours. Additionally, the neural network could be trained in a supervised manner, potentially enabling training on retrospectively undersampled motion scan data against known deformation fields calculated from fully sampled data. This presents a promising solution for further exploration of accelerating the motion scan acquisition.

Promising methods have also been recently proposed for motion-resolved quantitative MRI, particularly in cardiac imaging. Magnetic resonance multitasking for example, simultaneously resolves multiple time dimensions of undersampled image series via low-rank tensor imaging, allowing for motion-resolved parametric mapping with up to four time and/or contrast dimensions (\cite{christodoulouMagneticResonanceMultitasking2018}). This method could perform multi-parametric T1 and T2 mapping in the myocardium without requiring ECG information and under free-breathing conditions. The main challenge in implementing this approach within the context of the MRF T1-FF sequence lies in its intense computational demands. Our objective is to reconstruct five parametric maps (namely $T1_{H2O}$, FF, B0, B1, and $T1_{fat}$), while also accounting for the temporal dimension, all at high spatial resolution. High resolution is necessary to detect thin structures such as the diaphragm and intercostal muscles, and this must be achieved across a considerably large field of view. In our framework, akin to the approach introduced by \cite{cruzGeneralizedLowRank2022}, motion is compensated rather than resolved. Specifically, a reference respiratory phase is selected, and all other motion states are registered to this reference phase to generate motion-corrected parametric maps. However, it is noteworthy that by running the reconstruction step multiple times with different reference respiratory phases, it is feasible to obtain motion-resolved 3D parametric maps for $T1_{H2O}$ and FF, at the cost of increasing reconstruction time (Fig. S6).

After applying our reconstruction framework, the resulting motion-corrected images and parametric maps exhibited clear delineation of the liver and kidneys, showcasing distinct contrast between various anatomical regions such as blood vessels and lung interfaces for the liver, and calyces and medulla for the kidneys. Additionally, streaking artifacts were significantly reduced in the shoulder girdle and pectoral muscles. The diaphragm, previously scarcely visible in some subjects, consistently appeared in all subjects across several slices. The enhanced sharpness of anatomical regions facilitated precise delineation of regions of interest (ROIs) in the liver, kidneys, and respiratory muscles. In healthy volunteers, motion correction markedly improved the precision of measured variables. Specifically, we noted a systematic decrease in the standard variation of parameter distribution within different regions of interest.

In the liver, $T1_{H2O}$ and FF values measured with our approach aligned well with the literature. \cite{fellnerWaterFatSeparatedT12023} used a variable flip angle Dixon sequence acquired during breath-hold and found $T1_{H2O}$ of 873 ± 96 ms. \cite{shinNormalRangeHepatic2015} provided a normal range of FF in the liver between 0 and 0.08. In the kidney, $T1_{H2O}$ values were consistent across all our subjects but slightly higher than the T1 values obtained using a FISP MRF sequence (\cite{chenMRFingerprintingRapid2016}). However, to the best of our knowledge, specific $T1_{H2O}$ has never been assessed in the kidney, and hence the values are hard to benchmark to any other known methods for now. In the skeletal muscles, $T1_{H2O}$ and FF values were in the range of what has been measured in the lower limbs. For example, in \cite{martyWaterfatSeparationMR2021}, authors reported a mean $T1_{H2O}$ value of 1198 ± 50 ms in group of healthy volunteers. \cite{lopezkolkovskyMultiparametricAgingStudy2024} reported mean FF value between 0.034±0.010 and 0.068±0.019 in the posterior compartment of the leg in different groups of healthy volunteers from different age ranges. 

Compared to other muscles, the diaphragm and intercostal muscles exhibited greater inter- and intra-subject variability in both FF and $T1_{H2O}$ among the small group of healthy volunteers in our study. This variability could be attributed to their thin, layered structure, which makes delineation challenging and may result in values being affected by the partial volume effect depending on the orientation of the imaging plane. Since these muscles have not been previously evaluated using quantitative MRI, a larger number of subjects would need to undergo this protocol to establish reference values.

The clinical relevance of measuring $T1_{H2O}$ and FF has been extensively documented across various applications. In the field of neuromuscular imaging, for instance, FF is an established imaging biomarker of disease severity (\cite{carlierSkeletalMuscleQuantitative2016}), with demonstrated predictive capability for subsequent functional impairment (\cite{naardingMRIVastusLateralis2020}). Meanwhile, $T1_{H2O}$ has emerged as a promising biomarker of active muscle damages (\cite{hooijmansCompositionalFunctionalMRI2023}), exhibiting correlation with water T2 in inflammatory myopathies (\cite{martyWaterfatSeparationMR2021}). Despite their recognized utility, these variables have yet to be evaluated in respiratory muscles for patients, potentially offering insights into tissue structural changes associated with disease progression prior to functional impairment. Moreover, in liver imaging, FF has demonstrated greater sensitivity compared to biopsy for quantifying fat content in Non-Alcoholic Fatty Liver Disease (NAFLD), a condition affecting 5\% of the global population (\cite{noureddinUtilityMagneticResonance2013}). Furthermore, $T1_{H2O}$ has recently shown promise in differentiating between early-stage mild and severe NAFLD rodent models of liver disease (\cite{wanWaterSpecificMRI2022}).

Beside the relatively low number of subjects enrolled in this proof-of-concept study, our approach has several limitations. First, the total scan time currently ranges from 15 to 20 minutes, rendering it suitable for clinical research but nearly impractical for integration into routine clinical workflows in its current state. As mentioned earlier, the modularity of our approach provides various strategies for acceleration. Indeed, both scans could be independently optimized. For instance, deformation fields could be interpolated from lower resolution scans to decrease motion scan acquisition time. Partition undersampling techniques, as proposed in \cite{martyQuantitativeSkeletalMuscle2021}, could also be applied to the 3D MRF scan. Alternatively, optimizing the MRF pattern, as suggested by \cite{jordanAutomatedDesignPulse2021}, could enhance its sensitivity to the parameters of interest, thereby facilitating a reduction in scan time. Furthermore, in liver applications, T2* is typically assessed in addition to the other parameters to quantify iron content (\cite{henningerPracticalGuideQuantification2020}), thus it would be advantageous to incorporate it into the MRF sequence for this specific target.

\section{Conclusion}

In this study, we introduced a modular acquisition approach comprising a preliminary motion scan for estimating deformation fields, followed by an MRF scan for quantifying multiple parameters, accompanied by a reconstruction framework for iteratively reconstructing motion-corrected MRF singular volumes. This methodology significantly mitigated motion blurring and streaking artifacts in resulting 3D $T1_{H2O}$ and FF parametric maps with a spatial resolution of 1 x 1 x 5 mm$^3$, leading to marked improvements in precision. Consequently, accurate detection and delineation of moving structures was consistently achieved, with measured values aligning closely with those reported in the literature. These advancements pave the way for quantifying FF and $T1_{H2O}$ in regions previously challenging to assess due to motion, such as the respiratory muscles.

\section*{Acknowledgments}
Acknowledgments should be inserted at the end of the paper, before the
references, not as a footnote to the title. Use the unnumbered
Acknowledgements Head style for the Acknowledgments heading.

\section*{Declaration of Generative AI and AI-assisted technologies in the writing process} 
During the preparation of this work the authors used ChatGPT 3.5 in order to improve the writing style of this scientific paper. After using this service, the authors reviewed and edited the content as needed and take full responsibility for the content of the publication.

\section*{References}

%%Harvard
\bibliographystyle{unsrt} 
\bibliography{mocomrf}

\section*{Funding}

This work was supported by the French National Research Agency (ANR-20-CE19-0004) and the Association Institut de Myologie.

\end{document}